\documentstyle[prb,aps,epsf]{revtex}
\begin{document}
\title{Effect of disorder on the non-dissipative drag}
\author{John Baker$^{(a)}$, Giovanni Vignale$^{(b)}$ and A. G. Rojo$^{(a)}$ }
\address{ $^{(a)}$Department of Physics, University of Michigan, Ann Arbor, MI 48109-1120}
\address{$^{(b)}$Department of Physics, University of Missouri, Columbia, MO 65211}

\maketitle
\begin{abstract}
In this paper we consider the effect of disorder on the non--dissipative Coulomb 
drag between two mesoscopic metal rings at zero temperature.  Ring 1 has an 
Aharonov--Bohm flux present which creates a persistent current $J_{0}$.  Ring 2 
interacts with ring $1$ via the Coulomb potential and a drag current, $J_{d}$ is 
produced.  We show that this drag current persists with finite disorder in each 
ring, and that for small disorder, $J_{d}$ decreases with the square of the 
disorder amplitude.  
\end{abstract}
\section{Introduction}

Electron--electron (e-e)  interactions are responsible for a
multitude of \ fascinating effects in condensed matter. They play a leading
role in phenomena ranging from high temperature superconductivity and the
fractional quantum Hall effect, to Wigner crystalization, the Mott
transition and Coulomb gaps in disordered systems.  The effects of this
interaction on transport properties, however, are difficult to measure.\ \ A
new technique has\ recently proven effective in measuring the scattering
rates due to the Coulomb interaction directly\cite{gramila/prl}.

This technique is based on an earlier proposal by Pogrebinski\u{\i}\cite
{pogre,price0}. The prediction was that for two conducting systems separated
by an insulator (a semiconductor--insulator--semiconductor layer structure
in particular) there will be a drag of carriers in one film due to the
direct Coulomb interaction with the carriers in the other film. If layer $2$
is an ``open circuit'', and a current starts flowing  in layer $1,$ \ there
will be a momentum transfer to layer  $\ 2$ that will start sweeping carriers
to one end of the sample, and inducing a charge imbalance across the film.
The charge will continue to accumulate until the force of the resulting
electric field balances the frictional force of the interlayer scattering.
In the stationary state there will be an induced, or drag voltage $V_{D}$ in
layer $2$. \

There is a fundamental difference between transresistance and ordinary 
resistance insofar as the role of the Coulomb interaction is concerned.  
For a perfectly pure,
translationally invariant system, the Coulomb interaction cannot give rise
to resistance since the total current commutes with the Hamiltonian $H$.
This means that states with a finite current are stationary states of $H$
and will never decay, since the e-e interaction conserves not only the total
momentum but also the total current. (For electrons moving in a periodic
lattice, momentum and velocity are no longer proportional and the current
could in principle decay by the e--e interaction.)  If the layers are
coupled by the Coulomb interaction, the stationary states correspond to a
linear superposition of states in which the current is shared \ in different
amounts between layers: the total current within a given layer is not
conserved and can relax via the inter--layer interaction.

This mechanism of current degrading was studied in the pioneering experiment
of Gramila {\it et al.}\cite{gramila/prl} for GaAs layers embedded in AlGaAs
heterostructures. \ The separation between the layers was in the range $200$-%
$500$\AA . The coupling of electrons and holes and the coupling
between a two dimensional and a three dimensional system was also examined\cite
{solomon1}.

If we call $I$ the current circulating in layer $1$, the drag resistance (or
transresistance) is defined as
\[
\rho _{D}=\frac{V_{D}}{I}.
\]
Most of the experiments done so far indicate the vanishing of $\rho _{D}$ at 
zero temperature, something expected in the usual scattering theory of transport. 

\begin{figure}
\epsfxsize=4.0in
\vspace{5pt}
\hspace{3.cm}
\epsffile{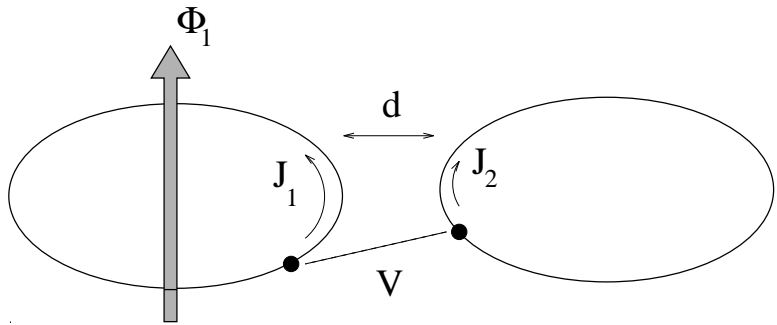}
\vspace{1.cm}
\caption{Schematic depiction of the non--dissipative drag setup. A persisent
current $J_1$ is induced in ring 1 by a Bohm--Aharonov flux. The Coulomb
interaction $V$ couples the charge fluctuations and generates a current
in the second ring}
\label{plot:rings}
 \end{figure}

The possibility of a drag effect at zero temperature was considered
by Rojo and Mahan\cite{rojo}, who considered two coupled \ mesoscopic\cite
{mesoscopics} rings that can individually sustain persistent currents, see 
Figure~(\ref{plot:rings}). 
%
The mechanism giving rise to drag in a non--dissipative system is also based on
the inter--ring or inter--layer Coulomb interaction, the difference with the
dissipative case being the coupling between real or virtual interactions.
One geometry in which this effect comes to life is two collinear rings of
perimeter $L$, with a Bohm--Aharonov flux, $\Phi _{1}$, threading {\em only one%
} of the rings (which we will call ring one). This is of course a difficult
geometry to attain experimentally, but has the advantage of making the
analysis more transparent. Two coplanar rings also show the same effect\cite
{rojo}. If the rings are uncoupled in the sense that the Coulomb
interaction is zero between electrons in different rings, and the electrons
are non--interacting within the rings, a persistent current $%
J_{0}=-cdE/d\Phi _{1}=ev_{F}/L$ will circulate in ring one\cite{buttiker}.
If the Coulomb interaction between rings is turned on, the Coulomb
interaction induces coherent charge fluctuations between the rings, and the
net effect is that ring two acquires a finite persistent current. The
magnitude of the \ persistent drag current $J_{D}$ can be computed by
treating the modification of the ground state energy in second order
perturbation theory $\Delta E_{0}^{(2)}$, and evaluating

\begin{equation}
\label{eq:persist0}
J_{D}=-e\left. \frac{d\Delta E_{0}^{(2)}}{d\Phi _{2}}\right| _{\Phi _{2}=0},
\end{equation}
with $\Phi _{2}$ an auxiliary flux treading ring two that we remove after
computing the above derivative. 

The question of the effect of disorder on persistent currents remains
controversial. Since our project involves calculating the effect of
disorder on an induced persistent current, we expect our results 
to shed some light on this issue. 
For an isolated pure ring the persistent current is $J_0=ev_F/L$ with 
$L$ the perimeter of the ring and $v_F$ the Fermi velocity. The most
immediate effect of disorder is to introduce a mean free path $\ell$.
One expects disorder to decrease the persistent current, and qualitative
arguments indicate that it is decreased by a factor $\ell/L$: $J_0 \rightarrow
ev_F/L (\ell/L)$. Our results indicate on firmer theoretical grounds that
a similar argument can be used for the drag persistent current. 

In this paper we outline our detailed studies of the effect of disorder on
non-dissipative drag using both analytic and numerical methods. 
\section{General remarks  on the non--dissipative Drag}

The zero drag current can be finite only if quantum coherence, or
entanglement, between the wave functions of the two systems is
established. In this situation, the meaningful description of the dynamics
of the combined system involves a single wave function, which distinghuishes
from ordinary dissipative drag, a case in which one has scattering between
two incoherently coupled systems.  Figure~(\ref{plot:rings}) is a schematic
 illustration of this coherent coupling mechanism.

\begin{figure}
\epsfxsize=3.in
\vspace{5pt}
\hspace{3.cm}
\epsffile{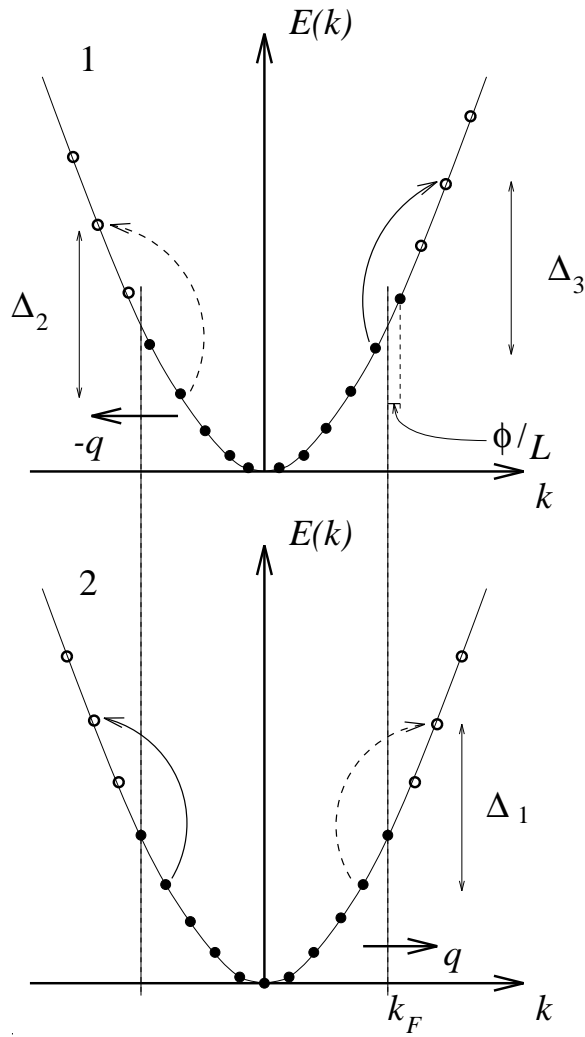}
\vspace{1.cm}
\caption{Schematic depiction of the non--dissipative drag mechanism. 
A Bohm--Aharonov flux $\Phi$ applied in ring 1 displaces the $k$ 
values of the  free one--particle states by $\phi/L$, with $\phi 2\pi\Phi/\phi_0$,
$\phi_0$ being the flux quantum. The unperturbed energy dispersions are therefore
asymmetric in ring 1 and symmetric in ring 2. When the interaction is turned on,
it creates 
virtual excitations of momentum $-q$ in ring 1 and momentum $q$ in ring 2 
(shown by the dashed arrows). The amplitude of this excitation is $V(q)/(\Delta_1
+\Delta_2)$.
 These excitations are not cancelled by
those of reversed momentum (shown by the continuous arrows), the amplitude 
of which is $V(q)/(\Delta_1
+\Delta_3)$ because of the asymmetry in the spectrum of 1. As a result there
is a persistent current in ring 2 proportional to $|V(q)|^2\phi/L$ from the 
terms indicated in this figure.
}
\label{basic_mec}
 \end{figure}

 We consider first two one-dimensional systems. 
 Assume that, in the absence
 of the Coulomb coupling,  system $1$ carries a finite equilibrium
 current, which could in principle be established by an Aharonov-Bohm
 flux threading system  $1$ only. 
 If system $2$ is a one dimensional wire of perimeter $2 \pi L$, 
 the mesoscopic nature of the zero drag current can be  proven
by the following analysis. 
 
 Let $\Psi_0$ be the  ground state of the combined system. 
This wave function involves the coordinates of both systems. Let us consider
system $2$ as a closed ring geometry, and designate
the coordinates of the particles in this subsystem as angular variables
$\theta _i$, with $i=1,\cdots,N_2$, and $N_2$ being the number  
of particles at system 2. The kinetic component of the Hamiltonian of system
$2$ can then be written as
\begin{equation}
H_K^{(2)}=-{\hbar ^2\over 2 mL^2} \sum_{i=1}^{N_2} 
{\partial ^2\over \partial \theta_i^2} \; .
\end{equation}

Consider the modified wave function $\Psi' $ constructed by applying a ``boost",
or gauge transformation, on the coordinates of system $2$:
\begin{equation}
\Psi' = U(\alpha) \Psi_0\equiv \exp(i\alpha\sum_{i=1}^{N_2}\theta_i)\Psi_0,
\end{equation}
with $\alpha $ a parameter. By the variational theorem $E'=\langle    \Psi'|
H | \Psi'\rangle \geq E_0$, with $H$ the Hamiltonian of 
the combined system, and $E_0$ the total energy. On the other hand, explicit evaluation of
$E'$ gives
\begin{equation}
E'=E_0+{\hbar^ 2\over 2mL^2} N_2 \alpha^2 
-{h \over e} \langle    \Psi_0 | \hat{J}_{\rm Tot}^{(2)} |\Psi_0  \rangle \, ,  
\end{equation}
with the current operator for  system $2$  given by
\begin{equation}
\hat{J}_{\rm Tot}^{(2)}={e\over 2 \pi m L^2} \sum_{i=1}^{N_2} 
i\hbar {\partial \over \partial \theta_i}\, .
\end{equation}

Due to the variational nature of the bound, the dragged current has to obey
the inequality:
\begin{equation}
J_{\rm drag } \equiv \langle \Psi_0 | \hat{J}_{\rm Tot}^{(2)} |\Psi_0
\rangle \leq
\alpha ^2{e \hbar \rho \over 2 \pi m L} \, ,
\label{jdrag}
\end{equation}
with $\rho $ the particle density. Equation (\ref{jdrag}) emphasizes the mesoscopic
nature of   the dragged current: in the limit of $L\rightarrow \infty$, 
$J_{\rm drag }\rightarrow 0$ with the same length dependence as the persistent 
current in mesoscopic rings, the value of which is $e v_F/L$ in the ballistic
regime. Note that the bound is valid for strictly one-dimensional systems. 

Having established a bound, one needs to show that there is indeed a finite dragged
current, and provide a quantitative estimate. We first present such a  calculation
treating the Coulomb interaction between the systems in second order perturbation
theory.
Consider two identical one-dimensional wires. Wire $1$ is threaded by a 
Aharonov-Bohm
flux $\phi_1$ (in units of the flux quantum). In order to evaluate the induced
current $J_2$,  we impose also a flux $\phi_2$ in system $2$, and  compute
 
\begin{equation}
J_2=-{e\ \over \hbar} \left.{ \partial E_0 \over \partial \phi_2}\right|_{\phi_2=0} \; .
\end{equation}

We neglect the Coulomb interaction within each wire, and consider the ballistic regime
(no impurities in either system).
In the absence of coupling, and for both  fluxes  $\phi_i < \pi/2$, the ground state 
consists of two Fermi systems with one particle energies $E^{(0)}_i ={\hbar ^2\over 2mL^2}(n_i-\phi_i)^2$,
and occupied levels for $n_i <n_F$ ($i=1,2$, and $n_F=N/2$, $N$ being the particle number
at each ring). Let $V(q)$ be the Fourier transform of the Coulomb
coupling, which
for wires separated a distance $d$ has the form $V(q)=(2 e^2/L) K_0(qd)$,
$K_0(x)$ 
being the zero-order Bessel function of imaginary argument. The second
order correction to the energy is then given by:
\begin{equation}
\Delta E _2=-{mL^2\over \hbar^2}\sum_{Q,n_1,n_2} {V^2({ Q\over L})\over Q}{f_{n_{1}}(1-f_{n_{1}+Q}) f_{n_{2}}(1-f_{n_{2}-Q})
\over (Q+n_1+\phi_1-n_2-\phi_2)} \; ,
\end{equation}
with $Q, n_1, n_2$ integers, and $f_m$  Fermi functions:  
$f_m =1$ if $|m| < n_F$, and zero otherwise.
The above sum is now evaluated  transforming the sum into integrals over
the continuum variables $q=Q/L$, $k_i=n_i/L$. Evaluating the 
integrals, and computing the derivative with respect to $\phi_2$, we
obtain
\begin{equation}
J_2=-{me^5\over\hbar^3} {1\over 2\pi^3} {I(k_Fd)\over k_FL} \phi_1 \;,
\end{equation}
with
$
I(k_Fd)=\int_0^{\infty} dq {q K_0^2(qd)  \over 4k_F^2 -q^2}
$. In the limit of large $k_Fd$, which corresponds to the interparticle
 distance being much smaller than the distance between the systems, we obtain
 \begin{equation}
 J_2\simeq J_0 {1\over (k_Fa_0)^2}{1\over (k_Fd)^2} \; ,
 \end{equation}
 with $J_0=-ev_F\phi_1/L$ being the persistent current carried by the otherwise
 uncoupled system $1$, and $a_0$ the Bohr radius. 
We have proven that there is an induced persistent current due to the
Coulomb interaction. We now ask ourselves about the induced effect if
system $2$ is made open, so that no current can circulate. 
In the transport  situation, a voltage will be induced. Here, we
 show that there is no voltage induced. We start with  a setup that,
in the absence of the flux in system $1$, is ``parity even". 
 By this we mean that the charge distribution in wire $2$ is 
symmetric around the center of the wire. We want to know if this symmetry is broken
by applying the flux in system $1$, an operation that breaks the
time reversal symmetry. Let us call $P$ and $T$ the parity and time 
reversal operators that interchange the ends of the wire. We want for example the induced dipole moment
in wire $2$, 
$x_2 =\langle \Psi_0 | \hat{x}_2 |\Psi_0\rangle$. The operator $
PT \hat{x}_2(PT)^{-1}=-\hat{x}_2$, while the wave function  is invariant
under $PT$, which implies $x_2=0$, hence there is no induced voltage.

\section{Disorder and the non--dissipative Drag}
In this section we outline our results on the effect of disorder in the 
non--dissipative drag.  In calculating the effects of disorder  we use the two ring geometry considered 
by Rojo and Mahan, see Figure~(\ref{plot:rings}), and calculate the second order Coulomb 
interaction between the 
conduction electrons in the two rings. The Coulomb potential is 

\[V \equiv\sum_{k}V_{k}\rho _{k,1}\rho_{-k,2}
=\int dx \int dx^{\prime} \rho_{1}(x) \rho_{2}(x^{\prime}) V(x-x^{\prime}),
\]   
With $\rho_{i}$ the charge density at ring $i$.  The second order correction to 
the ground state energy due to the Coulomb interaction is 

\begin{equation}
\label{eq:DE}
\Delta E= \sum_{n,n^{\prime}} \sum_{m,m^{\prime}}\frac{\arrowvert \sum_{k}V_{k} 
(\int   \psi_{n}e^{i\pi kx/L}\psi_{n^{\prime}}^{\ast}dx) (\int \psi_{m} e^{i\pi 
kx^{\prime}/L} 
\psi_{m^{\prime}}^{\ast} dx^{\prime}) f_{n^{\prime}}(1-f_{n}) 
f_{m^{\prime}}(1-f_{m}) \arrowvert^{2}}{E_{n}-E_{n^{\prime}} + 
E_{m}-E_{m^{\prime}}}, 
\end{equation}
where $\psi_{n}$ is an eigenstate in the presence of disorder.  From this 
expression for the energy shift we can calculate the drag current from 
Equation~(\ref{eq:persist0}). 

\subsection{Analytics}
\label{sec:analytics}	

In this section we estimate the effect of disorder on the non-dissipative drag current 
for the case in which disorder is present only in the ring on which the Bohm--Aharonov 
flux is applied. The driven ring (ring 2), on which the drag current circulates, will be taken 
as disorder-free. Momentum remains a good quantum number in ring 2  making the calculation more
tractable. The first order correction to the wave function is  given by

\begin{equation}
|\Psi_1\rangle = \sum_q V(q) \sum_k \sum_{\bar{\nu}}
{c^{\dagger}_{k+q} c_{k}|F_2\rangle |{\bar{\nu}}\rangle \langle{\bar{\nu}}| \rho_q |\psi_0^{(1)}\rangle
\over 
E_{k+q}-E_{k} +E_{\bar{\nu}}-E_{0,1}},
\end{equation}
where $E_{k}$ are the one-particle energies for the states of ring 2, and $|{\bar{\nu}}\rangle$ is a {\em many-body}
state of ring 1 with energy $E_{\bar{\nu}}$. The ground state of ring 1 is $|\psi_0^{(1)}\rangle$, and its energy is
$E_{0,1}$. Now, since we are neglecting interactions within each ring, the resulting equilibrium current in ring 2 is given
by
\begin{equation}
J_2={e\over L}\sum_q {\hbar q\over m} |V(q)|^2
\sum_{k}\sum_{\mu,\nu}{ f_k(1-f_{k+q})
f_\mu(1-f_\nu) |\langle \mu|e^{iqx}|
\nu\rangle|^2\over (E_{k+q}-E_k+E_{\nu}-E_{\mu})^2},
\label{drag_dis}
\end{equation}
where now $|\nu\rangle$ refers to the exact one-particle states with
energies $E_\nu$ corresponding to the  disordered Hamiltoninan in
ring 1. 
We can rewrite the above espression in terms of the spectral function
$S(q,\omega)$ defined
as
\begin{equation}
S(q,\omega)=\sum_{\mu,\nu}f_\mu(1-f_\nu)  |\langle \mu|e^{iqx}|\nu\rangle|^2
\delta\left(\omega -(E_{\nu}-E_{\mu})/\hbar\right).
\end{equation}

We will  consider the function $S(q,\omega)$ in the approximation in 
which the matrix element $|\langle \mu|\rho_q|\nu\rangle|$ is 
 given by the diffusive lorentzian\cite{imry}:

\begin{equation}
|\langle \mu|e^{iqx}|\nu\rangle|^2=
{1\over \pi \hbar N(0)}
{Dq^2 \over (Dq^2)^2 +(E_\mu-E_\nu)^2/\hbar^2},
\end{equation}
where $D$ is the difussion constant and $N(0)$ is the density of states of
the system. In this approximation we obtain that $S(q,\omega)$
is given by
\begin{equation}
S(q,\omega)= N(0) {\omega \, Dq^2 \over (Dq^2)^2 +\omega^2}.
\end{equation}

Before replacing this expression in Equation (\ref{drag_dis}) let us  recall
that there is a  flux
$\Phi$ threading ring 1 and therefore one expects $S(q,\omega)
\neq S(-q,\omega)$. We follow Ambegaokar and Eckern\cite{pcurr2} in including the
effect of the flux in the diffusive motion through the replacement:
\begin{equation}
Dq^2 \rightarrow D\bar{q}^2 \equiv D\left(q-\pi{\phi\over L}\right)^2,
\end{equation}
with $\phi$ being the flux in units  of the flux quantum.

The induced current will therefore be given by

\begin{equation}
J_2 ={e\over L} 
\sum_q  {\hbar q\over m} |V(q)|^2 \hbar D \bar{q}^2 N(0)
\sum_k \int d\omega {f_k (1-f_{k+q})\over (E_{k+q}-E_k+\hbar \omega)^2}
{\omega \over (D\bar{q}^2 + \omega^2)}.
\end{equation}

For small wavevectors ($q\ll k_F$) we have:

\begin{equation}
\sum_k {f_k (1-f_{k+q})\over (E_{k+q}-E_k+\hbar \omega)^2}
= {L\over 2\pi} {q\over \left({\hbar^2\over m}k_Fq +\hbar \omega\right)^2},
\end{equation}
and also, in the limit of $q\ell <1$, with $\ell$ being the mean free path:
\begin{equation}
\int_0^{\infty}d\omega {1\over \left({\hbar^2\over m}k_Fq +\hbar \omega\right)^2}{\omega
\over (D\bar{q}^2)^2+\omega^2}
\simeq { (\bar{q}\ell)\over (\hbar v_F)^2 q^2}.
\end{equation}
We are interested in the lowest order in $\phi$ for the induced current, 
which gives
\begin{equation}
J_2= {e\over 4\pi} N(0) {D\ell\over mv_F^2} {\phi\over L}
\sum_q q^2 V(q)^2,
\end{equation}
which we can now rewrite using $D=v_F \ell$ as
\begin{equation}
J_2\sim\left[\left({e v_F\over L}\right) \left( {\ell \over L}\right)\right]
\left[{\ell \over d}{N(0)(e^2/d)^2 C\over E_F}\right]\times \phi,
\label{anal_dis}
\end{equation}
where $C$ is a constant, 
\begin{equation}
C = \int_0^{\infty}dx x^2 K_0(x)^2=.308425
\end{equation}

The first term in square brackets in Equation (\ref{anal_dis}) corresponds to
a familiar expression for 
the persistent current in ring 1 in the presence of disorder. 
The value of terms in the second square bracket 
can be computed taking $N(0)=1/\Delta$, with $\Delta \sim 10K$ being
the level
spacing for a ring of $L\sim 1\mu m$, $E_F=2eV$, and a distance between rings
of $d=100$\AA. Note that this term contains the
product of two ratios: a small one given by 
$E_{\rm Coul}/E_F$, with $E_{\rm Coul}=e^2/d$, 
and a large one given by $E_{\rm Coul}/\Delta$. This gives a number of
order one, a result that  
probably overestimates the drag current, but serves as an indication that
the effects of disorder are not extreme. The second square bracket also
contains an additional ratio, the mean free path to the distance between
rings. This additional factor shows that the effects of disorder are 
stronger in the drag current from that in the driving ring. 
In order to test this results we performed numerical simulations, which
we present in the following sections.


\subsection{Numerical simulations}
\subsubsection{Perturbative treatment of the Coulomb interation}

In evaluating the drag current computationally we consider a discrete ring with 
$N$ lattice sites and $P<N$ electrons. We model disorder by placing a random 
disorder potential at each lattice site. The hamiltonian for an electron hopping 
between lattice sites in this ring is given by 

\[ 
H=t(\sum_{i=1}^{N}C^{\dagger}_{i}C_{i+1}e^{i\phi}+\sum_{i}C^{\dagger}_{i-1}C_{i}
e^ 
{-i\phi})+\sum_{n=1}^{N}W_{n}C_{n}^{\dagger}C_{n},
\]
where $\phi$ is the magnetic flux through the ring, $C^{\dagger}_{i}$ is the 
electron creation operator at site $i$ and $w_{n}$ is the disorder potential at 
site $n$.  For $N$ lattice sites, this gives an $N \times N$ hopping matrix.  

\begin{figure}
\epsfxsize=6.0in
\vspace{-5pt}
\hspace{3.cm}
\epsffile{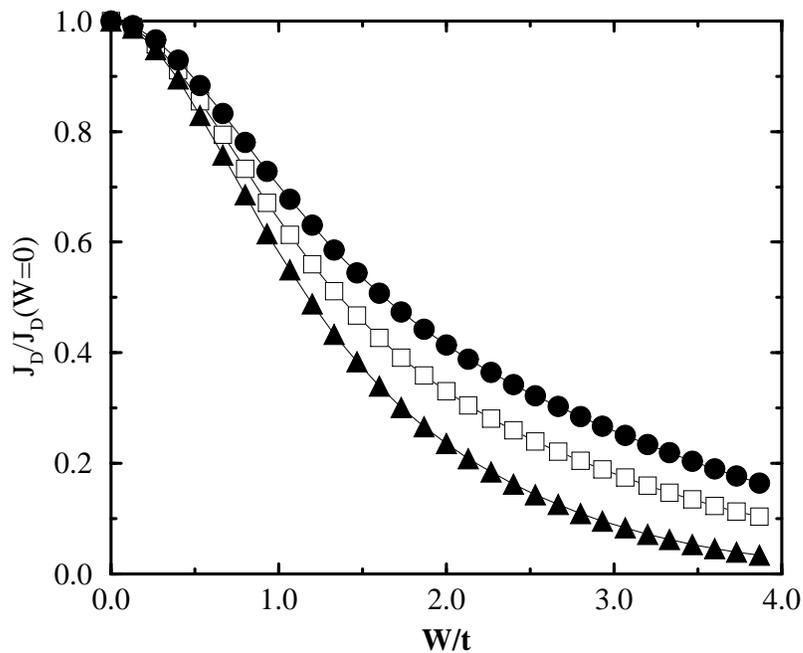}
\vspace{-1.cm}
\caption{Plot of Drag Current vs. Disorder Amplitude for a system of 10 lattice sites 
and 7 particles with disorder averaging.  We show three curves: $J_d/J_d(0)$ for one 
disordered ring 
(square symbols), $J_d/J_d(0)$ for both rings with disorder (triangles) and $J_o/J_o(0)$
(circles).}
\label{plot:JvsW}
    \end{figure}

In computing the energy shift for the two ring system we work with the x space 
representation of Equation~(\ref{eq:DE}), 
\begin{equation}
\label{eq: xspaceDE}
\Delta E= \sum_{n,n^{\prime}=1}^{N} \sum_{m,m^{\prime}=1}^{N} \frac{| 
\sum_{x} \sum_{x^{\prime}} V(x-x^{\prime}) 
<x|n><n^{\prime}|x><x^{\prime}|m><m^{\prime}|x^{\prime}>  
  f_{n^{\prime}}(1-f_{n}) 
f_{m^{\prime}}(1-f_{m}) |^{2}} 
{E_{n}-E_{n^{\prime}} + E_{m}-E_{m^{\prime}}}. 
\end{equation}
Here $x$ and $x^{\prime}$ denote discrete positions of the lattice sites in 
rings one and two respectively and the $|n>$'s and $E_{n}$'s are the 
eigenvectors and eigenvalues obtained numerically from the hopping matrix.  We 
obtain disorder averaging by evaluating $\Delta E$ with different realizations 
of the random disorder potentials, $W_{n}$, at values between $-W$ and $W$ where 
$W$ is the disorder amplitude.   The result of the computer simulations are shown 
in figure~(\ref{plot:JvsW}) for a system of 10 lattice sites and 7 particles. 
 The ratio of the drag current to its zero disorder value
$J_d/J_d(0)$ is plotted both for a system in which disorder is present in ring 2 only 
and for a system of two disordered rings.  The ratio $J_o/J_o(0)$ is also plotted.
For small  disorder amplitude, $J_{d} \propto W^{2}$.

\subsubsection{Non-perturbative treatment for very small rings by Lanczos method}

\begin{figure}
\epsfxsize=6.5in\hspace{8.cm}
\vspace{0.2cm}

\epsffile{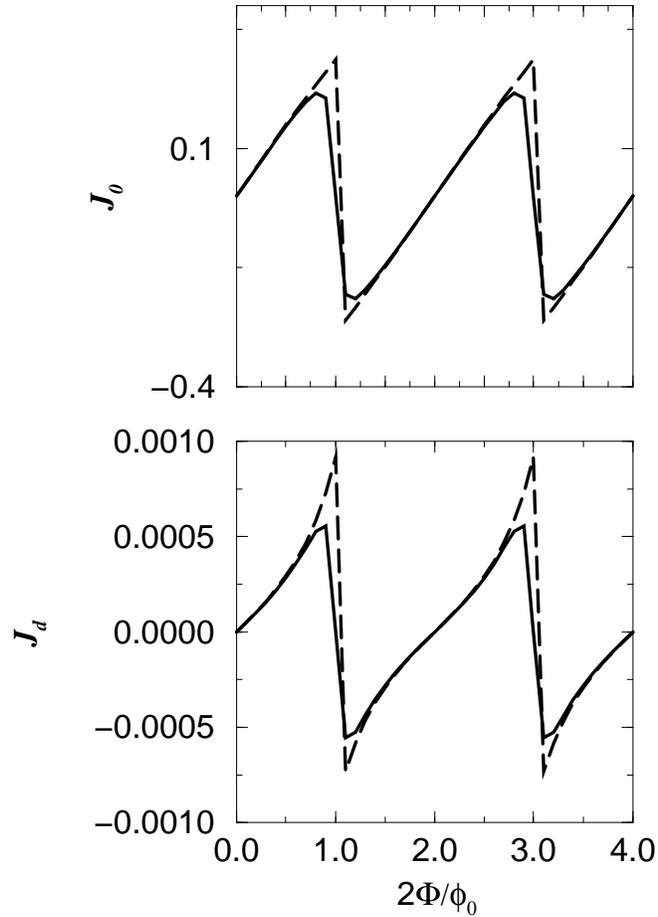}
\vspace{0.cm}
\caption{ Exact results for the persistent current and 
the drag current as a function of 
flux for two rings of six sites, each of them with two particles.
 We show curves for zero disorder as well as finite disorder.}
\label{plot:rings_lanzos}
 \end{figure}

In this section we present some exact results for small clusters. We use the
Lanczos method to diagonalize the problem, and obtain results
that are non--perturbative in the interaction. As a first illustration, 
Figure~(\ref{plot:rings_lanzos}) shows the persistent and drag currents both 
with and without disorder. The drag current follows the persistent current of 
ring 1 in its periodicity of one flux quantum as a function of the applied flux
through ring 1.

\begin{figure}
\epsfxsize=6.5in\hspace{8.cm}
\vspace{0.2cm}

\epsffile{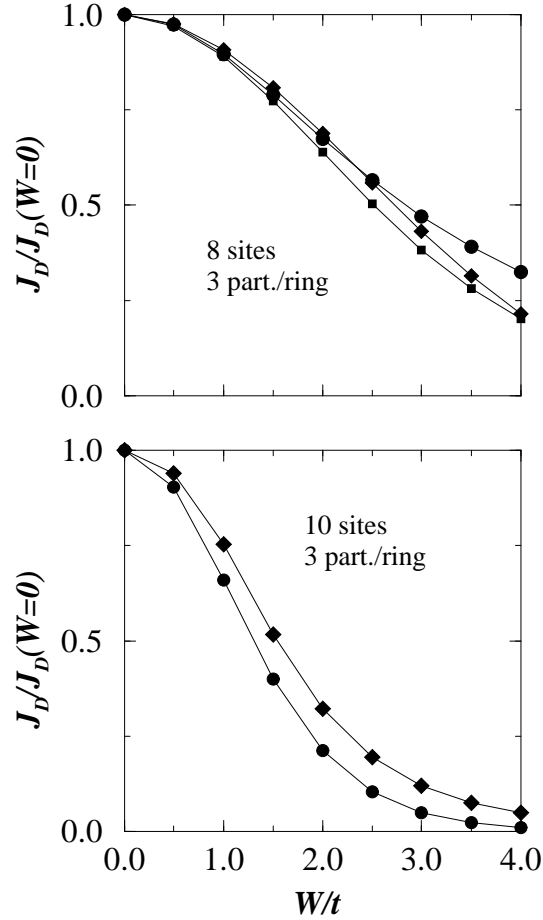}
\vspace{0.cm}
\caption{ Exact results for the drag current as a function of disorder amplitude
$W$ in units of the hopping matrix element for two rings interacting
via a delta function potential of amplitude $0.5t$. The flux in ring 1 is $\Phi_1=0.2\phi_0$}
\label{plot:rings_lanzos_2}
 \end{figure}

Figure~(\ref{plot:rings_lanzos_2}) shows the drag current for two systems of 
different sizes.  Note that the dependence with disorder is stronger for the larger 
system as expected from the factors of $\ell/L$ that appear in the analytical expressions
in section~\ref{sec:analytics}.
In conclusion we have established that the drag current remains finite for finite 
disorder.  We have shown by numerical simulations of finite clusters and
by analytical considerations that the effect of disorder on the drag current is more 
pronounced than the effect of disorder on the persistent current in a single ring.

\end{document}